\documentstyle[preprint,aps,floats]{revtex}
\begin{document}

\preprint{\tighten \vbox{\hbox{UT-preprint-0105}
  \hbox{hep-th/0110134}}}
\draft
\title{Renormalization Group Equations and the
Lifshitz Point\\
In Noncommutative Landau-Ginsburg Theory}
\author{Guang-Hong Chen and Yong-Shi Wu}
\address{Department of Physics, 
University of Utah,\\
Salt Lake City, Utah 84112, U.S.A.\\
\vspace{.5cm} 
{\tt ghchen@physics.utah.edu \\
     wu@physics.utah.edu}
}
\maketitle
{\tighten
\begin{abstract}
A one-loop renormalization group (RG) analysis 
is performed for noncommutative Landau-Ginsburg 
theory in an arbitrary dimension. We adopt a 
modern version of the Wilsonian RG approach, in 
which a shell integration in momentum space 
bypasses the potential IR singularities due to 
UV-IR mixing. The momentum-dependent trigonometric 
factors in interaction vertices, characteristic of 
noncommutative geometry, are marginal under RG 
transformations, and their marginality is preserved 
at one loop. A negative $\Theta$-dependent anomalous 
dimension is discovered as a novel effect of the 
UV-IR mixing. We also found a noncommutative 
Wilson-Fisher (NCWF) fixed point in less than four 
dimensions. At large noncommutativity, a momentum
space instability is induced by quantum fluctuations, 
and a consequential first-order phase transition is 
identified together with a Lifshitz point in the 
phase diagram. In the vicinity of the Lifshitz point, 
we introduce two critical exponents $\nu_m$ and 
$\beta_k$, whose values are determined to be $1/4$ 
and $1/2$, respectively, at mean-field level.
\end{abstract}}
\newpage

\section{Introduction}

Field theory on a noncommutative (NC) space, or simply 
noncommutative field theory (NCFT), has attracted 
much interest recently\footnote{For an incomplete 
list, see e.g. 
\cite{filk,kra,martin,bigatti,seiberg,hay,suss,gomis,chu,german,armoni,chenwu2}. For a more complete list, please see\cite{RMP}.}. 
The fact that NCFT arises in string/M(atrix) 
theory\cite{matrix,witten} suggests that space or 
spacetime noncommutativity should be a general feature 
of quantum gravity for generic points inside the moduli 
space of M-theory. As non-gravitational theory that has 
stringy features, studying NCFT is expected to shed new 
light on string/M theory. Moreover, as a natural deformation 
of usual quantum field theory, NCFT is of interest in its 
own right. In particular, NCFT is expected to be relevant 
to planar quantum Hall systems, since a charge in the 
lowest Landau level (LLL) in a strong perpendicular
magnetic field can be viewed as living in a noncommutative 
space: The guiding-center coordinates for the cyclotron 
motion of the charge in the LLL are known not to commute
\cite{girvin,read}. 

Of great importance both for quantum gravity and for 
the applications in condensed matter physics, one of the 
central issues about quantum dynamics of NCFT is to 
understand its low energy effective theory. At first glance, 
the answer to this question seems to be trivial according to 
the following reasoning: Low energies correspond to large 
distances. When the distance scale under consideration is 
much larger than the length scale given by the coordinate 
noncommutativity, the effects of the latter should be 
negligible. In other words, the effects of coordinate 
noncommutativity should be restricted to distances 
comparable to the noncommutative length scale. 
Therefore, it seems that noncommutative effects 
should disappear at sufficiently low energies. If this 
were true, one would expect that a strong magnetic 
field, which gives rise to a small noncommutative 
magnetic length, should have no effects in the 
large-distance physics of the charged particles in the 
LLL. From our experience with the quantum Hall systems 
we know that this is certainly incorrect. On one hand, 
a feature of the Laughlin wave function\cite{bob}, which
is crucial for its success, is that it is a wave function 
in the LLL. On the other hand, many people suspect that the 
problems with the Chern-Simons theory for half-filled 
LLL are related to the fact that it fails to properly 
incorporate the LLL condition\cite{read}. Thus, our 
experience with quantum Hall physics seems to indicate that 
coordinate noncommutativity should have characteristic 
observable effects in large distance phenomena. 

To see noncommutativity effects at large distances 
or at low energies requires a careful analysis of the 
effects of quantum fluctuations, involving 
renormalization and renormalization group (RG) flow. 
In studying one-loop renormalization of noncommutative 
scalar theory, Minwalla, Raamsdonk and Seiberg (MRS)
have discovered a novel effect\cite{seiberg}, called 
UV-IR mixing, which is characteristic of
coordinate noncommutativity. 
The physical essence of UV-IR mixing can be seen as
follows: Suppose we consider a noncommutative 
two dimensional Euclidean space, with 
\begin{equation}
 [x^\mu, x^\nu]=i\Theta^{\mu\nu} 
\hspace{1cm} \mu, \nu = 1,2,
\label{noncom}
\end{equation}
where the antisymmetric $\Theta^{\mu\nu}=\Theta
\varepsilon^{\mu\nu}$ ($\mu, \nu = 1,2$) contains a 
real parameter $\Theta$ of dimension length squared. 
Then, we have the uncertainty relation like, say,
$\Delta x^1 \Delta x^2 \geq \Theta $. For a wave 
packet on this noncommutative plane, if we make 
its size in $x^1$-direction small, then its size in 
$x^2$-direction will become big. So the UV effects 
in $x^1$-direction are entangled with the IR effects 
in $x^2$-direction. Conceptually, in our opinion, 
it must be the general notion of this UV-IR mixing 
that makes noncommutativity effects possible to 
manifest themselves at large distances. 

Let us examine the concrete effects of UV-IR mixing
found in Ref. \cite{seiberg}, to see whether it could 
help us understand the RG flow of NCFT to low energies. 
There they found IR singularities emerging as a 
trade-off of the UV divergences in loop momentum 
integration in non-planar diagrams. After removing 
the cut-off the corrected propagator will have an 
IR-singular piece even in the massive case. More
precisely, the non-planar part of the one-loop 
corrections to the $\phi$-propagator in 
$\phi^4$ theory in $D$ dimensions, is given by 
\begin{eqnarray}
\Sigma_{np} (p) &=& g \int^{\Lambda} 
\frac{d^D k}{(2\pi)^D} \frac{e^{ip\wedge k}}{k^2+m^2}
\nonumber \\
&=& \frac{g}{(4\pi)^{D/2}} \int_{0}^{\infty} 
\frac{d\alpha}{\alpha^{D/2}} \exp\{-\alpha m^2 
-\frac{p \circ p}{4\alpha}- \frac{1}
{\Lambda^2\alpha}\},
\label{np-prop}
\end{eqnarray}
where $\Lambda$ is a UV regulator, $g^2$ the coupling
constant and $p \circ q
= -p_{\mu}q_{\nu}\Theta_{\mu\lambda}\Theta^{\nu}_{\lambda}$.
This results in a corrected inverse propagator of 
the form\cite{gubser}
\begin{eqnarray}
\Gamma^{(2)}(p)=p^2+m^2+2g I_2(0)+g I_2(p),
\label{corr-prop}
\end{eqnarray}
with
\begin{eqnarray}
I_2(p) = (2\pi)^{-D/2} m^{\frac{D-2}{2}}
\left( p\circ p+ \frac{4}{\Lambda^2}\right)^{\frac{2-D}{4}}
 K_{\frac{D-2}{4}}\left(m
\sqrt{p\circ p+ \frac{4}{\Lambda^2}}\right),
\label{I-2}
\end{eqnarray}
$K_\nu (x)$ being a modified Bessel function.
It is easy to see that the limit $\Lambda \to 
\infty$ does not commute with the low momentum 
limit $p \to 0$. If one insists to take the limit 
$\Lambda \to \infty$ first, then $\Gamma^{(2)}(p)$
acquires an IR singularity that grows to infinity as
$p\to 0$. This seems to make a renormalization 
group analysis difficult to carry out.  

However, an RG analysis is essential for a good 
understanding of the phase structure and the low 
energy effective theory of NCFT, especially for 
potential applications in the quantum Hall systems.  
In this paper we attempt to present a Wilsonian 
RG analysis for NCFT, in the form recently 
proposed by Shankar\cite{shankar} in the context 
of condensed matter physics and by 
Polchinski\cite{polchinski} in the context of high 
energy physics.  We observe that in this modern RG 
approach to low energy effective theory, the 
momentum cut-off, $\Lambda$, that defines the theory 
is thought of as the momentum above which the 
dynamical degrees of freedom have been integrated 
out. Therefore one should {\it not} take the 
cut-off to infinity. Rather, it should be kept 
finite while we do RG transformations to eliminate 
degrees of freedom between $\Lambda$ and $\Lambda/s$ 
with $s>1$. Therefore, it is the integration in momentum 
space over the shell between $\Lambda/s$ and $\Lambda$ 
(with $s>1$) that is relevant to the RG transformation. 
This is a shell in momentum space with finite radius; 
obviously this shell integration bypasses the potential 
IR singularity found in Ref. \cite{seiberg}. It remains 
to see how noncommutativity effects manifest themselves 
at low energies (with increasing $s$). This is the 
subject of this paper.

For simplicity, in this paper, we treat a Euclidean
version of NC $\phi^4$ theory as a noncommutative 
Landau-Ginsburg (NCLG) theory. Namely, we explicitly
introduce temperature into the theory through the
``mass'' term: $m^2=a_2(T-T_c)$, where $a_2>0$ is a 
constant and $T_c$ is the critical temperature. Other 
coefficients in front of $\phi^n$ do not explicitly 
depend on temperature.

In performing RG analysis, a crucial issue is how 
to deal with the RG transformation of the momentum
dependent Moyal phase or trigonometric factors in 
the interaction vertices (see eq. (\ref{interaction}) 
below), which arise from coordinate noncommutativity 
(\ref{noncom}), or equivalently from the Moyal's 
star product (\ref{star}). One thought might be that 
when we scale momentum $k \to k'= sk$ 
under RG transformations, the noncommutativity 
parameters $\Theta^{\mu\nu}$ appearing in these 
factors, since they are of dimension of length
squared, should transform like $\Theta^{\mu\nu}
\to \Theta^{\prime \mu\nu}= s^{-2}\Theta^{\mu\nu}$,
and therefore should eventually flow to zero at 
low energy (with $s\to \infty$). This is equivalent
to renormalizing the NC-parameters in defining the
renormalized interaction vertices. However, this is 
in conflict with the notion of noncommutative 
geometry, i.e. with the idea that $\Theta_{\mu\nu}$ 
are among the parameters that describe geometry 
of the space (or spacetime), which are unaffected 
by any non-gravitational dynamics. Therefore, 
after RG transformation the interaction vertices 
should still be written in the form of a star product 
of fields with the {\it same} $\Theta_{\mu\nu}$, 
rather than the re-scaled ones. As we will see 
below, this leads to a classification of relevance 
of operators according to the original star product 
rather than one that eventually flows to the usual 
product. This will have an important implication to 
the RG flow of interactions at low energy: Namely, 
interactions in the effective low-energy action are 
always those with an invariable star product; they 
can never flow to the interactions with the usual 
product. In this sense we say that the Moyal 
trigonometric factors in interaction vertices in 
NCFT is marginal at tree level, and remains marginal 
at one loop level with the same $\Theta^{\mu\nu}$. 
 
The main results of the present paper reveal a new 
manifestation of the general notion of UV-IR mixing: 
Namely, in or near four Euclidean dimensions, 
non-planar contributions from the noncommutative 
interaction vertices at one loop induce a 
renormalization of the quadratic kinetic term in the 
low energy effective action. This effect is distinct 
for NCFT, because it is well-known that in ordinary 
scalar theory in or near four dimensions there is no 
wave function renormalization at all at one loop. On 
one hand, the new effect indicates a momentum space 
instability for sufficiently large values of 
noncommutativity parameter $\Theta$. On the other hand, 
it gives rise to a {\em negative, and $\Theta$-dependent} 
critical exponent (or anomalous dimension) $\eta$ at 
the noncommutative Wilson-Fisher (NCWF) fixed point in 
less than four dimensions. This effect certainly makes 
noncommutativity visible at distances that are much 
larger than the noncommutative scale, an astonishing 
entanglement between the UV (at noncommutative scale) 
and the IR (at the scale of the correlation length). 
When the noncommutativity exceeds a critical value 
$\theta_c$, the large distance asymptotic behavior of 
the correlation function would change its form from a 
usual negative power law to a positive power law, this 
signals a phase transition. In momentum space, this 
phenomenon corresponds to a susceptibility $G(k)$ that 
changes from the $1/k^2$ to $1/k^4$ behavior, implying 
an instability in momentum space: The coefficient of 
the $k^2$ term for kinetic energy in the effective 
action becomes zero or negative, so that the dynamical 
behavior is dominated by the $k^4$ term, provided it 
has a positive coefficient to maintain the system stable. 
This instability leads to the appearance of a Lifshitz 
(multi-critical) point in the phase diagram of the 
system, as is known in the theory of phase transitions 
and critical phenomena\cite{Fla-Hag,Horn}. In this way, 
we have shown a quantum fluctuation induced modulated 
phase in the NCLG theory. The scaling behavior and 
corresponding critical exponents in the vicinity of the 
Lifshitz point are also studied. 

The present paper is the long version of a previous short 
letter\cite{chenwu}, presenting more calculations and 
adding more detailed discussion on the Lifshitz point. 
Before our letter, Ref. \cite{gubser} had considered 
the possibility of a transition to a non-uniform phase 
in the NCLG theory, and conjectured the appearance of a 
Lifshitz point in the phase diagram. Our reasoning for
the transition to a modulated phase is different from 
theirs, and our demonstration of the existence of the
Lifhsitz point is based on a solid RG analysis. After 
our letter, Ref. \cite{italy} confirmed our way of 
doing renormalization in NCFT, and also addressed the
issue of the Wilsonian RG equation. We will discuss
the differences between these papers and ours in the 
final section. 
   
The paper is organized as follows: In Sec. II, we 
present an introduction to RG analysis, {\it \`a la} 
Shankar and Polchinski, for the quadratic terms in 
the NCLG theory, determining the RG transformations 
for the field and the mass parameter at tree level. 
This section is elementary and experts in condensed 
matter RG can just skip it. Then in Sec. III we study 
the peculiarities due to noncommutativity in classifying 
the relevance of quartic perturbations at tree level. 
One loop RG equations for quadratic and quartic vertices 
are discussed in Sec. IV; in particular, the anomalous 
dimension of the field and rescaling of the momentum 
dependence of the Moyal phase factor from the star 
product are further discussed at one-loop level. The 
subsequent section, Sec. V, is devoted to exploring the 
consequences of this novel one-loop anomalous dimension 
(wave function renormalization) and to discussing the 
NCWF fixed point. In Sec. VI, we present an analysis 
of the fluctuation induced first-order phase transition 
and the Lifshitz point. The final section (Sec. VII) is 
devoted to conclusions and discussions.

\section{Fixed Point and Quadratic Terms}

We start with the NCLG theory in $D$ dimensions: 
\begin{equation}
  \label{action}
S=-\int d^{D}x[\frac{1}{2}(\partial_{\mu}\phi)^2
+\frac{1}{2}m^2\phi^{2}
+\frac{g}{4!}\phi*\phi*\phi*\phi],
\end{equation}
where the star product of two functions is defined as
\begin{equation}
\label{star}
(f*g)(x)=e^{\frac{i}{2}\Theta^{\mu\nu}\partial^{y}_{\mu}
\partial^{z}_{\nu}} f(y)g(z)|_{y=z=x},
\end{equation}
with $\Theta^{\mu\nu}$ real and antisymmetric. For 
convenience, we choose the
noncommutativity matrix to be 
\begin{equation}
\label{theta}
\Theta^{12}=\Theta^{34}=\Theta
\end{equation}
and other independent components being zero. 
The generalization to anisotropic cases is 
straightforward.

In the following, we present a Wilsonian RG 
analysis in the modern formulation recently 
proposed by Shankar\cite{shankar} and 
Polchinski\cite{polchinski}. The basic 
procedure is as follows: One starts with a 
free system with momentum cut-off $\Lambda$, 
and identifies the RG scaling laws so as to 
make the free action invariant (a fixed 
point of RG transformations in the low energy 
regime). After doing so, we turn on the 
interactions, consider them as perturbations 
to the fixed-point action, and examine the 
interaction vertices both at the tree and 
one-loop levels. According to their behavior 
under RG scaling, the interactions can be 
classified as relevant, irrelevant, or marginal 
ones. Only the relevant and marginally relevant 
interactions can possibly change the fixed point, 
i.e. change the low energy physics. So {\em in the 
low energy effective field theory, we keep only 
the relevant and marginally relevant interactions}. 
Now let us apply this general procedure to the 
$D$ dimensional action (\ref{action}).

We rewrite the free field action $S_0$ in 
momentum space as
\begin{equation}
\label{free}
S_0=-\frac{1}{2}\int_{|k|<\Lambda}
\frac{d^Dk}{(2\pi)^D}k^2\phi(-k)\phi(k).
\end{equation}
First separate the field variable $\phi(k)$ 
into slow modes $\phi_s(k)$ and fast modes 
$\phi_f(k)$:
\begin{eqnarray}
\label{modes}
\phi_s(k)&=&\phi(k),\qquad |k|<\frac{\Lambda}{s}, \\
\phi_f(k)&=&\phi(k), \qquad \frac{\Lambda}{s}\le 
|k|\le\Lambda,
\end{eqnarray}
where $s$ is a number greater than unity. 
Then let us integrate out the fast modes in the partition
function $Z=\int{\cal D}\phi e^{S}$;
since 
they are quadratic in the action, the result is a constant 
and does not contribute to the critical properties. We 
are interested in the resulting action for the slow modes; 
after rescaling $k'=sk$ to make the cut-off back to 
$\Lambda$, it becomes 
\begin{eqnarray}
\label{freeprime}
S^{\prime}_0(\phi_s)&=&-\frac{1}{2}
\int_{|k|<\frac{\Lambda}{s}}\frac{d^Dk}{(2\pi)^D}
k^2\phi_{s}(-k)\phi_{s}(k), \\  \nonumber
&=&-\frac{1}{2}s^{-D-2}\int_{|k^{\prime}|<\Lambda}
\frac{d^Dk^{\prime}}{(2\pi)^D}k^{\prime 2}
\phi_{s}(-\frac{k^{\prime}}{s})\phi_{s}
(\frac{k^{\prime}}{s}).
\end{eqnarray}
To make this action invariant, we redefine 
the field variable as
\begin{equation}
\label{newfield}
\phi^{\prime}(k^{\prime})=s^{-\frac{D}{2}-1}\phi_s
(\frac{k^{\prime}}{s})=s^{-\frac{D}{2}-1}\phi_s(k).
\end{equation}
With the definition (\ref{newfield}),  
eq. (\ref{freeprime}) reads
\begin{equation}
\label{invariant}
S^{\prime}_0(\phi_s)= 
S_0(\phi^{\prime})
=S^*,
\end{equation}
which tells us that the free action $S_0$ 
is a fixed point under the following
RG transformations
\begin{equation}
\label{rgtrans}
k^{\prime}=sk, \hspace{1.0cm} \phi^{\prime}
(k^{\prime})=s^{-\frac{D}{2}-1}\phi_s (k).
\end{equation}

Having the fixed point in hand, we can 
immediately proceed to investigate the 
relevance of the mass term in the action 
(\ref{action}). After integrating out 
the fast modes we have
\begin{eqnarray} 
\label{s2}
S^{\prime}_2(\phi_s)&=&-\frac{1}{2}m^2\int_{|k|<\frac{\Lambda}{s}}
\frac{d^Dk}{(2\pi)^D}\phi_{s}(-k)\phi_{s}(k), 
\\ \nonumber
&=&-\frac{1}{2}m^2s^{-D}\int_{|k^{\prime}|<\Lambda}
\frac{d^Dk^{\prime}}{(2\pi)^D} \phi_{s}
(-\frac{k^{\prime}}{s})\phi_{s}(\frac{k^{\prime}}{s}), 
\\ \nonumber
&=&-\frac{1}{2}m^2s^{2}\int_{|k^{\prime}|<\Lambda}
\frac{d^Dk^{\prime}}{(2\pi)^D}\phi^{\prime}(-k^{\prime})
\phi^{\prime}(k^{\prime}).
\end{eqnarray}
{}From the last equality sign, we derive the 
following scaling relation for the mass parameter
\begin{equation}
\label{mass}
r^{\prime}\equiv m^{\prime 2}=s^2r\equiv s^2m^2.
\end{equation}
Recalling that $s>1$, the mass term is enhanced under 
RG transformations (\ref{rgtrans}). Namely it flows 
away from zero -- its value at the RG fixed point.  
We conclude that the mass term is relevant under 
the RG flow to the low energy regime. This finishes 
the RG analysis for the quadratic actions at tree 
level. Note that noncommutativity does not show up 
in the quadratic action, consequently above analysis 
is identical to those in ordinary space.

\section{Classification of relevance for 
Quartic Perturbations: Tree Level}

In this section, we study the quartic 
perturbations in the action (\ref{action}):
\begin{equation}
\label{s4}
S_4=-\frac{1}{4!}\int_{\Lambda}\phi(4)
\phi(3)\phi(2)\phi(1)\, u(4321),
\end{equation}
where the 
interaction function $u(4321)$ is given by
\begin{eqnarray}
\label{interaction}
u(4321)&=& gI(4321)\equiv
\frac{g}{3}\biggl[
\cos(\frac{k_1\wedge k_2}{2})
\cos(\frac{k_3 \wedge k_4}{2})\\ \nonumber
&+& \cos(\frac{k_1\wedge k_3}{2})
\cos(\frac{k_2\wedge k_4}{2}) 
+ \cos(\frac{k_1\wedge k_4}{2})
\cos(\frac{k_2\wedge k_3}{2})\biggr],
\end{eqnarray}
with $k_1\wedge k_2=\Theta^{\mu\nu}k_{1\mu}k_{2\nu}$. 
We write the interaction function in this way 
so as to make the vertex totally symmetric; we 
will see that this representation has some 
advantage in counting the symmetry factors.
The short-hand notation of the integral in 
eq. (\ref{s4}) is understood as 
\begin{equation}
\label{int}
\int_{\Lambda}=\int_{|k_i|<\Lambda}\prod_{i=1}^{4}
\frac{d^Dk_i}{(2\pi)^{D}}\delta^{(D)}(k_1+k_2+k_3+k_4).
\end{equation}

Unlike the quadratic part discussed in the preceding 
section, the slow modes and fast modes get mixed up 
in the quartic case. To perform the mode elimination, 
in general we have to go back to the perturbative 
expansion. Here we use the cumulant expansion widely 
used in the study of critical phenomena\cite{gold}. 

To see how the cumulant expansion works, we 
cast $S_4(\phi)$ into $S_4(\phi_s, \phi_f)$
and perform path integral over fast modes,  
to obtain the partition function in the form
\begin{eqnarray}
\label{partition}
Z&=&\int{\cal D}\phi_s{\cal D}\phi_f 
e^{S_0(\phi_s)}e^{S_0(\phi_f)}e^{S_4(\phi_s,\phi_f)}, 
\\ \nonumber
&=&\int{\cal D}\phi_s e^{S^{r}(\phi_s)},
\end{eqnarray}
where the reduced action $S^{r}(\phi_s)$ is  
defined as
\begin{eqnarray}
\label{reduced}
e^{S^{r}(\phi_s)}&\equiv& e^{S_0(\phi_s)}
\frac{\int{\cal D}\phi_f e^{S_0(\phi_f)}
e^{S_4(\phi_s,\phi_f)}}{\int{\cal D}
\phi_f e^{S_0(\phi_f)}}, \\ \nonumber
&=&e^{S_0(\phi_s)}\biggl<e^{S_4(\phi_s,\phi_f)}
\biggr>_{0f}, \\ \nonumber
&=&e^{S_0(\phi_s)+S_4^{\prime}(\phi_s)}.  
\end{eqnarray}
Here we have used the fact that the functional 
integral in the denominator is only a constant, 
so we can freely drop it; and $<...>_{0f}$ 
denotes the average over the fast modes weighted 
by the free action of fast modes. 

The basic idea of the cumulant expansion is to 
relate the mean of exponential to the exponential 
of means, namely
\begin{equation}
\label{cumulant}
\biggl<e^{S_4(\phi_s,\phi_f)}\biggr>_{0f}
=\exp\biggl[\biggl<S_4(\phi_s,\phi_f)\biggr>_{0f}
+\frac{1}{2}\biggl(\biggl<S_4^2(\phi_s,\phi_f)
\biggr>_{0f}-\biggl<S_4(\phi_s,\phi_f)\biggr>^2_{0f}
\biggr)+\cdot\cdot\cdot\biggr].
\end{equation}
In this way, we get a perturbative series for 
$S_4^{\prime}(\phi_s)$:
\begin{equation}
\label{series}
S_4^{\prime}(\phi_s)=\biggl<S_4(\phi_s,\phi_f)
\biggr>_{0f}+\frac{1}{2}\biggl(
\biggl<S_4^2(\phi_s,\phi_f)\biggr>_{0f}
-\biggl<S_4(\phi_s,\phi_f)\biggr>^2_{0f}\biggr)+\cdots .
\end{equation}

Now let us apply this perturbative expansion to 
The quartic interactions in NC $\phi^{4}$ theory. 
Consider first the leading term in 
$S_4^{\prime}(\phi_s)$:
\begin{equation}
\label{s4s}
\biggl<S_4(\phi_s,\phi_f)\biggr>_{0f}
=-\frac{1}{4!}\biggl<\int
(\phi_s+\phi_f)_4(\phi_s+\phi_f)_3
(\phi_s+\phi_f)_2 (\phi_s+\phi_f)_1u(4321)\biggr>_{0f}.
\end{equation}
The right side contains $16$ terms in total, which 
include $8$ terms each containing an odd number of 
fast modes, one term purely fast modes, one term
purely slow modes, and the remaining six terms two 
fast and two slow modes. It is easy to see that the 
terms involving an odd number of fast modes vanish 
due to symmetry of the integral. The term involving 
only fast modes gives us a constant anyway, so we 
are not interested in it. In the following, we only 
need to concentrate on the last two classes, i.e. 
the term only involving slow modes and the six 
terms involving two fast and two slow modes.

The term with only slow modes gives us the tree 
level contribution to $S_4^{\prime}$, since the 
average over fast modes gives us unity:
\begin{eqnarray}
\label{s4tree}
S_{4,tree}^{\prime}&=&-\frac{1}{4!}
\int_{|k|<\frac{\Lambda}{s}}\phi_{4s}
\phi_{3s}\phi_{2s}\phi_{1s}u(4321), \\ \nonumber 
&=&-\frac{1}{4!}s^{-3D}\int_{|k^{\prime}|<\Lambda}
\phi_{4s}(\frac{k^{\prime}_4}{s})\phi_{3s}
(\frac{k^{\prime}_3}{s})\phi_{2s}(\frac{k^{\prime}_2}{s})
\phi_{1s}(\frac{k^{\prime}_1}{s})u(\frac{4^{\prime}}{s}
\frac{3^{\prime}}{s}\frac{2^{\prime}}{s}\frac{1^{\prime}}{s}),
\\ \nonumber
&=&-\frac{1}{4!}s^{4-D}\int_{|k^{\prime}|<\Lambda}
\phi^{\prime}_{4s}(k^{\prime})\phi^{\prime}_{3s}(k^{\prime})
\phi^{\prime}_{2s}(k^{\prime})\phi^{\prime}_{1s}(k^{\prime})
u(\frac{4^{\prime}}{s}\frac{3^{\prime}}{s}\frac{2^{\prime}}{s}
\frac{1^{\prime}}{s}), \\ \nonumber
&=&-\frac{1}{4!}\int_{|k^{\prime}|<\Lambda}
\phi^{\prime}_{4s}(k^{\prime})\phi^{\prime}_{3s}(k^{\prime})
\phi^{\prime}_{2s}(k^{\prime})\phi^{\prime}_{1s}(k^{\prime})
u^{\prime}(4^{\prime}3^{\prime}2^{\prime}1^{\prime}).
\end{eqnarray}
In the second and third line, the RG transformation 
(\ref{rgtrans}) have been used. Therefore, at tree 
level, we derive the scaling for the interaction 
function $u(4321)$ as follows
\begin{equation}
\label{treeuscale}
u^{\prime}(4^{\prime}3^{\prime}
2^{\prime}1^{\prime})=s^{4-D}u(\frac{4^{\prime}}{s}
\frac{3^{\prime}}{s}\frac{2^{\prime}}{s}
\frac{1^{\prime}}{s})=s^{4-D}u(4321).
\end{equation}

If one follows the usual procedure in the RG 
analysis on ordinary space, he/she would expand the 
cosine factors in the interaction function 
(\ref{interaction}) in powers of momenta, and apply 
the above RG transformation to each term. At tree 
level this would lead to
\begin{equation}
\label{tay}
u^{\prime}(4^{\prime}3^{\prime} 
2^{\prime}1^{\prime})=
s^{4-D}g-\frac{gs^{2-D}}{12}
[k^{\prime}_1\wedge k^{\prime}_2+
k^{\prime}_3\wedge k^{\prime}_4]+\cdot\cdot\cdot.
\end{equation}
Here the ellipsis represents terms with $s^{-D}$ 
or higher negative powers of $s$, which have no 
chance to be relevant in any dimension $D$. Thus,
except for the $\Theta$-independent constant term, 
all the higher-order $\Theta$-dependent terms would 
be {\em irrelevant} under the RG transformation, 
so that noncommutativity at tree level would be 
irrelevant to the low energy physics. Certainly 
this is in conflict with the general notion of
UV-IR mixing. 

What was wrong in the above procedure is apparently 
that one should {\it not} have expanded the Moyal 
factor $I(1'2'3'4')$ into power series of momenta. 
We should bear in mind that this momentum dependent 
factor originates from the star product, i.e. from 
coordinate noncommutativity, an intrinsic geometric 
feature of the NC space that ought to be respected 
by any non-gravitational dynamics. In other words, 
{\em dynamics of non-gravitational systems ought 
to  respect NC geometry}. In our opinion, it is 
this constraint from NC geometry on dynamics that 
constitutes the main new feature for the RG 
analysis on an NC space. 

So the problem with the expansion (\ref{tay}) is 
just that it did not respect the star product in 
the NC space, that coherently organizes infinitely 
many higher-order derivative terms. Though each of 
them may behave like an irrelevant operator in 
ordinary sense, however their coherent sum may 
give rise to non-trivial effects. That a coherent 
structure in the interactions may change the 
relevance of an interaction operator has been 
known to happen in ordinary field theory. One 
famous example is the $1+1$ dimension sine-Gordon 
model $S=\int d^2x [\frac{1}{2}\partial_{\mu}
\Phi\partial_{\mu}\Phi +g\cos(\beta\Phi)]$. 
If we expand the interaction term as an infinite 
Taylor series, then a simple scaling analysis tells 
us that every term is a relevant operator. However, 
it is well-known that this is not quite right. 
Actually the coherent structure $\cos(\beta\Phi)$ 
has scaling dimension $\Delta=\frac{\beta^2}{4\pi}$, 
so whether it is relevant depends on whether 
$\Delta>2$ (irrelevant) or $\Delta<2$ (relevant), 
and $\Delta=2$ (marginal). Similarly in NCFT, it 
is the structure of the Moyal's star product, 
dictated by the intrinsic NC geometry, that neatly 
and coherently organizes infinitely many higher 
derivative terms, modifying the classification of 
relevance of operators in NCFT.  

With these points in mind, one should define the 
renormalized interactions in terms of star products 
and classify their relevance, not that of ordinary 
interactions. Indeed, at least at one loop, it has 
been shown\cite{seiberg} that counterterms have 
the same star product structure. Thus, the 
operators allowed to appear in the Wilsonian 
effective action must be always of the the 
form of a star product with the same $\Theta$ 
parameter. The most general quartic terms are
always of the form of Eq. ({\ref{interaction}), 
with the coefficient $g$ becoming a function 
of momenta: $u(4321)= g(4321)\, I(4321)$.  We 
should classify the relevance of the 
interaction operator by expanding only the 
coefficient $g(k_1,k_2,k_3,k_4)$ into powers 
of momenta, while keeping the Moyal factor 
$I(4321)$ intact as if it is {\it marginal}.

To summarize, the difference of our treatments 
from the usual RG analysis in the commutative 
case is that we keep the star product structure 
of the quartic interaction intact, and apply RG 
transformations only to the coefficient function 
$g(k_1,k_2,k_3,k_4)$. Then define their relevance, 
irrelevance etc as usual. Thus at tree level 
$g^{\prime}=s^{4-D}g$, namely, it is a relevant 
operator in less than four dimension, it is 
irrelevant in higher than four dimension and it 
is marginal in the four dimension. In next
section, we will study one-loop effects, and see
that the ``marginality'' of the Moyal factor will be 
preserved at one loop level; i.e. it is indeed
protected from quantum fluctuations intrinsically 
by NC geometry.

\section{One Loop RG analysis}
\subsection{One loop corrections to the quadratic action}
Now let us proceed to examine the 
terms with two fast and two slow modes in 
Eq. (\ref{s4s}). Of these six terms, one typical 
term is of the form 
\begin{equation}
\label{vanishterms}
\int_{|k|<\frac{\Lambda}{s}}\frac{d^D k_1}{(2\pi)^D}
\int_{|k|<\frac{\Lambda}{s}}\frac{d^D k_2}{(2\pi)^D}
\int_{\frac{\Lambda}{s}<|k|<\Lambda}
\frac{d^D k_3}{(2\pi)^D}\int_{\frac{\Lambda}{s}
<|k|<\Lambda}\frac{d^D k_4}{(2\pi)^D}\phi_{4f}
(k_4)\phi_{3f}(k_3)\phi_{2s}(k_2)\phi_{1s}(k_1).
\end{equation}
 Since the free action of 
fast modes is of the form $\int_{\frac{\Lambda}{s}
<|k|<\Lambda} k^2\phi_f(-k)\phi_f(k)$ and the interaction
function $u(4321)$ is symmetrized, so all the six terms 
give the same contribution. Again, using the symmetry in 
the integral over the fast modes, we should take
\begin{equation}
\label{condition}
k_3=-k_4=p,\hspace{1.0cm} k_1=-k_2=k
\end{equation} 
to yield nonzero result. So we are led to the 
following simple result
\begin{eqnarray}
\label{qua}
-6&\times&\frac{g}{4!3}
\int_{|k|<\frac{\Lambda}{s}}\frac{d^Dk}{(2\pi)^D}
\phi_s(-k)\phi_s(k)\int_{\frac{\Lambda}{s}
<|k|<\Lambda} \frac{d^Dp}{(2\pi)^D}
\biggl<\phi_f(-p)\phi_f(p)\biggr>_{0f}
[2+\cos(k\wedge p)], \\ \nonumber
&=&-\frac{1}{2}\int_{|k|<\frac{\Lambda}{s}}
\frac{d^Dk}{(2\pi)^D}\phi_s(-k)\phi_s(k)\Gamma_2(k),
\end{eqnarray}
where the self-energy $\Gamma_2(k)$ is defined by
\begin{equation}
\label{self}
\Gamma_2(k)=\frac{g}{6}\int_{\frac{\Lambda}{s}<|p|<\Lambda}\frac{d^Dp}
{(2\pi)^D}\biggl[\frac{2}{p^2}
+\frac{\cos(k\wedge p)}{p^2}\biggr].
\end{equation}
In passing, we would like to remark that the above result can be 
summarized in a Feynman diagram, the tadpole diagram in Fig.1, the 
only difference from the familiar rule in quantum field theory being  
that the loop momentum only takes its value in a thin momentum shell 
$[\frac{\Lambda}{s},\Lambda]$. In addition, since the mass term is a 
relevant operator, we can also start from the full quadratic action to
 do the perturbation theory, the only change is the propagator, namely,
$1/p^2$ is changed to $1/(p^2+m^2)$. This change will be used later to
get the RG equation for the mass term. 

The planar contribution to the quadratic action from the first 
term in eq. (\ref{self}) is
\begin{eqnarray}
\label{quar}
-\frac{g}{12}\int_{|k|<\frac{\Lambda}{s}}
\frac{d^Dk}{(2\pi)^D}\phi_s(-k)
\phi_s(k)K_D\Lambda^{D-2}
(1-\frac{1}{s}), \\ \nonumber
=-\frac{g}{12}\int_{|k^{\prime}|<\Lambda}
\frac{d^Dk^{\prime}}{(2\pi)^D}\phi^{\prime}_s
(-k^{\prime})\phi^{\prime}_s(k^{\prime})
K_D\Lambda^{D-2}s^2(1-\frac{1}{s}),
\end{eqnarray}
where the integration variable $p$ in the 
integrand has been replaced by its upper 
limit $\Lambda$, because the thickness, 
$\Lambda(1-\frac{1}{s})$, of the integration 
shell in momentum space is very small. 
Moreover, $K_D=S_D/(2\pi)^D$ and $S_D$ is 
the surface area of a unit sphere in $D$ 
dimensions. Clearly this gives a correction 
to the mass term (\ref{s2}).

Now let us calculate non-planar contribution 
to the quadratic action from the second term 
in eq. (\ref{self}):
\begin{equation}
\label{nonplane} 
-\frac{g}{12}\int_{|k|<\frac{\Lambda}{s}}
\frac{d^Dk}{(2\pi)^D}\phi_s(-k)\phi_s(k)I(k).
\end{equation}
where $I(k)$ has been defined as
\begin{equation}
\label{integ}
I(k)=\int_{\frac{\Lambda}{s}<|p|<\Lambda}
\frac{d^Dp}{(2\pi)^D}\frac{e^{ik\wedge p}}{p^2}.
\end{equation}
To calculate this integral, we take 
$\Theta^{\mu\nu}$ to be of the form of 
eq. (\ref{theta}), then $k\wedge p$ is calculated as
\begin{equation}
\label{kwedgep}
k\wedge p=\Theta\sum_{i=0}^{\frac{D}{2}-1}
(k_{2i+1}p_{2i+2}-k_{2i+2}p_{2i+1}).
\end{equation}
To be more specific, 
we take $n=2$ as a example and then generalize it to the 
more general case.
We write 4-dimensional vectors $k$ and $p$ as
\begin{eqnarray}
\label{dec}
P=\vec{p}_{\parallel}+\vec{p}_{\perp},\hspace{1.0cm}
\vec{p}_{\parallel}=p_1\vec{e}_1+p_2\vec{e}_2, 
\hspace{1.0cm} 
\vec{p}_{\perp}=p_3\vec{e}_3+p_4\vec{e}_4,\\
\nonumber
K=\vec{k}_{\parallel}+\vec{k}_{\perp},
\hspace{1.0cm}
\vec{k}_{\parallel}=k_1\vec{e}_1+k_2\vec{e}_2, 
\hspace{1.0cm} \vec{k}_{\perp}
=k_3\vec{e}_3+k_4\vec{e}_4.
\end{eqnarray}
Then we choose polar coordinates in 12-plane 
and 34-plane with the azimuthal angles 
$\varphi_1$ and $\varphi_2$, respectively. 
The four dimensional integral (\ref{integ}) can be 
cast into
\begin{equation}
\label{polar}
I(k)=\frac{1}{(2\pi)^4}\int dp_{\parallel}
p_{\parallel}d\varphi_1\int dp_{\perp}p_{\perp}
d\varphi_2\frac{e^{i\Theta p_{\parallel}
k_{\parallel}\sin\varphi_1}e^{i\Theta p_{\perp}
k_{\perp}\sin\varphi_2}}
{p^2_{\parallel}+p^2_{\perp}}.
\end{equation}
After performing the integral over azimuthal 
angles, we get
\begin{equation}
\label{bessel}
I(k)=\frac{(2\pi)^2}{(2\pi)^4}\int dp_{\parallel}
p_{\parallel}\int dp_{\perp}p_{\perp}
\frac{J_0(\Theta p_{\parallel}k_{\parallel})
J_0(\Theta p_{\perp}k_{\perp})}
{p^2_{\parallel}+p^2_{\perp}}.
\end{equation}
Noting that the integral over $p$ has the shell 
constraint, namely, we have
\begin{equation}
\label{constraint}
\biggl(\frac{\Lambda}{s}\biggr)^2\le 
p^2_{\parallel}+p^2_{\perp}\le\Lambda^2.
\end{equation}
Introducing polar coordinate on the 
$(p_{\parallel},p_{\perp})-$ plane with 
the azimuthal angle $\varphi$, we get
\begin{eqnarray}
\label{shellinteg}
I(k)&=\frac{1}{4\pi^2}&\int_{\Lambda/s}^{\Lambda}
 dp p\int_{0}^{\pi/2}d\varphi\sin\varphi\cos\varphi
 J_0(\Theta k_{\parallel}p\cos\varphi)
 J_0(\Theta k_{\perp}p\sin\varphi),
\\ \nonumber
&=&\frac{\Lambda^2}{8\pi^2}(1-\frac{1}{s})
\int_{0}^{\pi/2}d\varphi\sin2\varphi
J_0(\Theta k_{\parallel}\Lambda\cos\varphi)
J_0(\Theta k_{\perp}\Lambda\sin\varphi).
\end{eqnarray}
By expanding 
the Bessel function up to second order, i.e. 
\begin{equation}
\label{taylor}
J_0(x)=1-\frac{1}{4}x^2+O(x^4),
\end{equation}
and calculate the integral over the angular variable
,we get
\begin{equation}
\label{Ik2}
I(k)=K_4\Lambda^{2}(1-\frac{1}{s})
(1-\frac{1}{8}\Theta^2\Lambda^2 k^2).
\end{equation}
The above procedure can be easily generalized to 
the case of $D=2n$, the result turns out to be
\begin{equation}
\label{Ik}
I(k)=K_D\Lambda^{D-2}(1-\frac{1}{s})
(1-\frac{1}{8}\Theta^2\Lambda^2 k^2).
\end{equation}
The constant term on the right side contributes
to mass renormalization. Combining it with the
contribution from eq. (\ref{quar}), we get the 
one-loop correction to the scaling relation
(\ref{mass}):
\begin{equation}
\label{oneloopmass}
r^{\prime}=s^2r + \frac{g K_D\Lambda^{D}}
{2(\Lambda^2+m^2)}(s^2-s).
\end{equation}
Setting $s=1+t$, where $t$ is infinitesimal, 
and taking derivative with respect to $t$, 
then we get the one-loop $\beta$-function 
for the mass term: 
\begin{equation}
\label{betaofm}
\frac{dr}{dt}=2r+\frac{u}{2}K_D(1-r), 
\end{equation}
where $u=g\Lambda^{4-D}$ has been used.

Moreover, the appearance of a momentum 
dependent term in eq. (\ref{Ik}) is a 
distinct feature due to coordinate 
noncommutativity, since in contrast, in 
the usual commutative case (with $\Theta=0$)
$I(k)$ does not depend on momentum at all. 
However, only a quadratic momentum dependence 
is marginally relevant, so we need Taylor expand 
eq. (\ref{taylor}) only up to second order at
this moment. By substituting the second term 
of eq. (\ref{Ik}) into eq. (\ref{nonplane}), 
we see that the noncommutative $\phi^4$ 
interactions induce a renormalization
of the kinetic term at one loop. Using field 
theory terminology, it gives rise to {\it wave 
function renormalization} for the $\phi$-field, 
that explicitly {\it depends on noncommutativity}. 
This is one of our key observations in this paper. 
With this term, the Gaussian fixed point action 
$S_0$ is modified to
\begin{eqnarray}
\label{s0mod}
S^{\prime}_0&=&-\frac{1}{2}\biggl[1-\frac{u}
{48} K_D(\Theta\Lambda^2)^2 t \biggr]
\int^{\frac{\Lambda}{s}}_0 k^2\phi_s(-k)\phi_s(k)
\end{eqnarray}
Using $1+ \gamma t \approx s^{\gamma}$, and 
introducing a dimensionless $\theta=\Theta\Lambda^2$ 
for noncommutativity, we see that the $\phi$-field 
acquires an anomalous dimension, modifying 
eq. (\ref{rgtrans}) to 
\begin{equation}
\label{scalinglaw}
\phi^{\prime}(k^{\prime})=s^{-\frac{D+2-
\gamma(u,\theta)}{2}}\phi(k)
\end{equation}
with the one-loop $\gamma$ depending on 
noncommutativity:
\begin{equation}
\label{anomalous}
\gamma(u,\theta)=-\frac{1}{48}uK_D\theta^2.
\end{equation}
This is a novel result, because in ordinary 
Landau-Ginsburg theory the anomalous dimension 
vanishes at one loop. Our result is a 
consequence of the UV/IR mixing in NCLGM. 
Also unusual is that the this anomalous dimension 
is {\it negative}. It may significantly affect 
the stability of the symmetric phase, as will 
become clear in the next section.

\subsection{One loop corrections to the quartic interaction}

We have seen that the leading term in the 
cumulant expansion produces the one-loop 
effects for the quadratic terms, but provides 
only tree level information on interaction 
vertices. To deal with the one-loop effects
for vertices, we need to invoke the sub-leading 
terms in eq. (\ref{series}):
\begin{equation}
\label{sublead}
\frac{1}{2}\biggl[\biggl<S_4^2(\phi_s,\phi_f)
\biggr>_{0f} -\biggl<S_4(\phi_s,\phi_f)
\biggr>^2_{0f}\biggr].
\end{equation}
By repeating the arguments used before for the 
renormalization of the quadratic terms, we 
see that the contributions to the vertices 
come from the terms of the following form:
\begin{equation}
\label{typical}
 \int_{|k|<\frac{\Lambda}{s}}\phi_s(4)
\phi_s(3)\phi_s(2)\phi_s(1)
\biggl<\int_{\frac{\Lambda}{s}<|k|<\Lambda}
\phi_f(8)\phi_f(7)\phi_f(6)\phi_f(5)
{\cal P}(k_8,k_7,\cdots,k_1)\biggr>_{0f},
\end{equation}
where ${\cal P}(k_8,k_7,\cdots, k_1)$ is 
composed of trigonometric functions, whose 
specific form will be given later. Repeating 
the same reasoning in the preceding section, 
we know that the non-vanishing contributions 
can only come from the Feynman diagrams 
shown in Fig. 2.

Since the interaction function has been symmetrized, 
we can easily count the symmetry factor for each 
diagram to be $1/2$. If we assume the momentum flow 
into the diagram is $Q$, the loop momentum is $q$, 
then the function ${\cal P}$ for the first diagram 
is given by \cite{jabbari}
\begin{eqnarray}
\label{verP}
&{\cal P}&(p_1, p_2, p_3, p_4;q)=
\frac{g^2}{9}\cos(\frac{p_1\wedge p_2}{2})\cos(\frac{p_3
\wedge p_4}{2})\biggl[1+\cos q\wedge(p_1+p_2)\biggr] \\ \nonumber
&+&\frac{g^2}{18}\cos(\frac{p_3\wedge p_4}{2})
\biggl[\cos(\frac{p_1\wedge p_2}{2}+q\wedge p_1)
+\cos(\frac{p_1\wedge p_2}{2}-q\wedge p_2)\biggr] \\ \nonumber
&+&\frac{g^2}{18}\cos(\frac{p_1\wedge p_2}{2})
\biggl[\cos(\frac{p_3\wedge p_4}{2}-q\wedge p_3)
+\cos(\frac{p_3\wedge p_4}{2}+q\wedge p_4)\biggr] \\ \nonumber
&+&\frac{g^2}{36}\biggl[\cos[\frac{p_1\wedge p_2-p_3
\wedge p_4}{2}+q\wedge(p_1+p_3)]+
\cos[\frac{p_1\wedge p_2+p_3\wedge p_4}{2}
+q\wedge(p_1+p_4)]\biggr].
\end{eqnarray}  
By proper permutation of the external momenta, 
we can get the ${\cal P}$ function for the 
other two diagrams. Noting that $p_i$ ($i=1,2,3,4$)  
are slow mode momenta and the loop integral is only 
within the momentum shell $[\frac{\Lambda}{s},\Lambda]$, 
we can put the external momenta to zero whenever 
they wedge the loop momentum \cite{gold,wilson}. 
This approximation greatly simplifies the above 
${\cal P}$ function, and leads to
\begin{equation}
\label{Psim}
{\cal P}(p_1,p_2,p_3,p_4)=\frac{g^2}{2}
\cos(\frac{p_1\wedge p_2}{2})
\cos(\frac{p_3\wedge p_4}{2}).
\end{equation}
After permutation on the external momenta, we 
can get a similar structure for the other two 
diagrams in Fig. 2. Thus the one loop corrections 
to the quartic interactions are given by
\begin{eqnarray}
\label{Cquar}
& &\frac{g^2}{2}\int_{|k|<\frac{\Lambda}{s}}I(4321)
\phi_s(4)\phi_s(3)\phi_s(2)\phi_s(1)
\int_{\frac{\Lambda}{s}}^{\Lambda}\frac{d^{D}p}
{(2\pi)^{D}}\frac{1}{(p^2+r)^2}, \\ \nonumber
&=&\frac{g^2}{2}\frac{K_D\Lambda^{D}}{(\Lambda^2+r)^2}
(1-\frac{1}{s})\int_{|k|<\frac{\Lambda}{s}}I(4321)\phi_s(4)
\phi_s(3)\phi_s(2)\phi_s(1), \\ \nonumber
&=&\frac{g^2}{2}\frac{K_D\Lambda^{D}}{(\Lambda^2+r)^2}s^{4-D}
(1-\frac{1}{s})\int_{|k|<\Lambda}I(\frac{4^{\prime}}{s}
\frac{3^{\prime}}{s}\frac{2^{\prime}}{s}\frac{1^{\prime}}{s})
\phi^{\prime}(4^{\prime})\phi^{\prime}(3^{\prime})\phi^{\prime}
(2^{\prime})\phi^{\prime}(1^{\prime}).
\end{eqnarray}
where $I(4321)=3u(4321)/g$ is the Moyal factor 
defined by the square bracket in eq. 
(\ref{interaction}). Now we come to the question 
of how to scale the $u(4321)$ again. To get 
the one loop RG for the quartic vertex, we have
to correctly handle the scaling for the trigonometric
factor $I(4321)$. At tree level, we have maintained 
that this factor is untouched when we do rescaling.
We need to check the consistency of this treatment 
when the loop effects are included. To get the one 
loop RG, we set $s=1+t$ with $t$ is small and 
positive. We consider one term in $I(4321)$, say
\begin{equation}
\cos(\frac{k_1^{\prime}\wedge k_2^{\prime}}{2s^2})
\cos(\frac{k_3^{\prime}\wedge k_4^{\prime}}{2s^2})=
\cos(\frac{k_1^{\prime}\wedge k_2^{\prime}}{2}
-tk_1^{\prime}\wedge k_2^{\prime})\cos(\frac{k_3^
{\prime}\wedge k_4^{\prime}}{2}-tk_3^{\prime}
\wedge k_4^{\prime})
\end{equation}
The small $t$ expansion gives
\begin{eqnarray}
\label{expan}
\cos(\frac{k_1^{\prime}\wedge k_2^{\prime}}{2})
\cos(\frac{k_3^{\prime}\wedge k_4^{\prime}}{2})
&+&tk_3^{\prime}\wedge k_4^{\prime}\cos(\frac{k_1^{\prime}
\wedge k_2^{\prime}}{2})\sin(\frac{k_3^{\prime}
\wedge k_4^{\prime}}{2}) \\ \nonumber
&+&tk_1^{\prime}\wedge k_2^{\prime}\cos(\frac{k_3^{\prime}
\wedge k_4^{\prime}}{2})\sin(\frac{k_1^{\prime}
\wedge k_2^{\prime}}{2}).
\end{eqnarray}
Therefore, after rescaling, some extra terms are 
generated. However, all extra terms are irrelevant 
in the sense of RG, since they all contain higher 
order derivatives (e.g. $k_1\wedge k_2$ etc.). We 
can discard them in the low energy effective action. 
Thus the consistency check within one loop is passed; 
namely, we do not need to scale the momentum
dependence in the star product. In passing we would
like to emphasize that the irrelevant terms in 
eq. (\ref{expan}) is due to a small $t$ expansion, 
not a small external momentum expansion as in 
eq. (\ref{tay}). After this discussion on the
star product, the RG equation of the quartic 
vertex is readily derived; the result is actually 
the same as its commutative counterpart:
\begin{equation}
\label{betau}
\frac{du}{dt}=(4-D)u-\frac{3}{2}K_Du^2.
\end{equation} 
However, we should notice that the NC effective 
theory is different from its commutative 
counterpart, because we need one more parameter,
the noncommutative parameter $\Theta$, to specify 
the star product in the interactions, unless the 
low energy effective theory is a free theory, 
as in the cases with $D\ge 4$ as discussed 
in the next section. 

\section{Stability Analysis of the Gaussian 
Fixed Point}

In this section, we will concentrate on the 
physical implications of our RG analysis 
presented in the preceding sections. Since 
the physics critically depends on the dimensionality
of space, we will discuss the $D\ge 4$ and 
$D<4$ cases separately.

\subsection{Above the critical dimension: $D\ge 4$}

According to eq. (\ref{betau}) the upper 
critical dimension in NCLG remains to be 
$D=4$. When $D>4$, the quartic coupling  
$u$ flows to zero, so it is irrelevant. 
Thus, it is the unique, trivial Gaussian 
fixed point, $u^{*}=0=r^{*}$, that controls 
the IR asymptotically free low-energy 
behavior, with the same sets of critical 
exponents as usual in mean field theory: 
$\nu=1/2$ and $\eta=\gamma(u^{*},\theta)=0$. 

However, this is not the whole story, when 
we consider approaching to the critical 
point $r=0$. In fact, the modified scaling 
law (\ref{scalinglaw}) gives a two-point 
correlation function behaving like 
\begin{equation}
\label{correlation}
\langle\phi(x)\phi(0)\rangle\sim\frac{1}
{x^{D-2 +\gamma}}.
\end{equation}
Due to the minus sign in (\ref{anomalous}), 
for very large noncommutative parameter
$\theta$, the above correlation function 
at large distance does not become zero, 
which signals an instability in the system. 
The critical value, $\theta_c$, is given 
by the condition
\begin{equation}
\label{condition2}
u\theta^2_c=\frac{48(D-2)}{K_D}.
\end{equation}
More precisely, the parameter space for 
the NCLG is now three-dimensional, 
described by $(u,r,\theta)$. The condition 
(\ref{condition2}) gives us a surface 
in parameter space. To access the Gaussian 
fixed point, we have to fine-tune the 
parameter $\theta$ to make $\theta < 
\theta_c(u)$.
Of course, the closer it gets to the fixed 
point, the less important is the condition 
(\ref{condition2}), since $\theta_c$ is 
pushed to infinity when the Gaussian fixed 
point is reached.

In the critical dimension $D=4$, the 
one-loop non-zero anomalous dimension 
(\ref{anomalous}) is expected to modify 
the logarithmic corrections to the scaling 
laws at criticality.

\subsection{Below the critical dimension:
physics in $4-\varepsilon$ dimension}
According to eq. (\ref{betau}), if the 
dimension is slightly lower than four, 
the Gaussian fixed point becomes unstable, 
and we have a new IR stable fixed 
point, the noncommutative counterpart of the 
Wilson-Fisher (NCWF) fixed point. Besides the 
noncommutativity parameter $\theta$, its 
position in $(r,u)$-space for small 
$\varepsilon\equiv 4-D$ is the same as in 
ordinary Landau-Ginsburg model (OLGM):
\begin{equation}
\label{wf}
u^{*}=\frac{16\pi^2}{3}\varepsilon,\hspace{1.0cm}
 r^{*}=-\frac{1}{6}\varepsilon.
\end{equation}
At this fixed point, the critical exponent 
$\nu$ is unchanged: 
$\nu=\frac{1}{2}+\frac{\varepsilon}{12}$, but the 
one-loop critical exponent $\eta$ becomes
non-vanishing:
\begin{equation}
\label{wfeta}
\eta= \gamma (u^{*}, \theta)=
-\frac{\varepsilon\theta^2}{72}.
\end{equation}
This result is characteristic of the NCLG. 
We would like to stress three important 
aspects of the critical exponent (\ref{wfeta}): 
(1) It starts at order of $\varepsilon$, while
in OLGM it starts at order of $\varepsilon^2$. 
(2) It is {\it negative}, while positive in 
OLGM. (3) It looks like non-universal because 
of its dependence on the dimensionless 
noncommutativity parameter $\theta$. 
However, in the present case, the NCWF 
fixed point had better be viewed as a 
line of fixed points labeled by $\theta$. 
Since $\theta$ originates in the microscopic
sector of the system, its appearance in the 
macroscopic critical exponent is a genuine 
manifestation of {\it UV/IR mixing}, namely, 
the fingerprint of a "high-energy" parameter
in low-energy phenomena.

To see how the anomalous dimension 
(\ref{wfeta}) affects the stability of 
the NCWF fixed point, let us examine the
two-point correlation function 
\begin{equation}
\label{corrscaling}
\langle\phi(x)\phi(0)\rangle\sim\frac{1}
{x^{2-\varepsilon (1+\theta^2/72)}}.
\end{equation}
The criterion to maintain the stability 
of the NCWF fixed point is that the above
correlation function should tend to zero 
at large distances. Now we have an 
additional parameter $\theta$ as a new 
knob to tune the system. If it is too 
large, the correlation function may become 
finite or even divergent. The critical 
value is given by
\begin{equation}
\label{criticaltheta}
\theta_c=12/\sqrt{\varepsilon},
\end{equation}
which depends only on $\varepsilon$. 
Therefore, if $\theta<\theta_c$, the NCWF 
fixed point is stable. On the contrary, 
if $\theta>\theta_c$ the NCWF fixed point 
will no longer be stable. This is 
reflected in the phase diagram, Fig. 3.

This situation is similar to previous RG 
analyses in the literature for one 
dimensional and three dimensional interacting 
fermion systems. There the RG analysis was 
used to show a similar instability for the 
Fermi liquid fixed point. To get the picture 
of the Luttinger liquid (in 1$d$) and BCS 
superconductivity (in 3$d$), one had to 
determine nonperturbatively the underlying 
physics for the new phase. Here to gain 
knowledge of the new phase for $\theta>\theta_c$, 
we also need extra efforts. This will be 
presented in the next section.

\section{First order phase transition at large $\theta$}

In the previous section, based on a one loop RG 
analysis we found that the critical exponent 
$\eta$ takes a negative value $-\varepsilon\theta^2/72$,
which depends on space dimensionality and 
noncommutativity parameters. In this section, we 
are going to demonstrate that system experiences 
a first order phase transition to a modulated
phase at the large $\theta_c=12/\sqrt{\varepsilon}$ 
and critical temperature. 
 
At first sight, one may wonder whether the appearance 
of a negative $\eta$ is a bad news for NCFT. Indeed, in 
the pure commutative $\phi^4$ theory, one can prove 
\cite{parasi,CG,fla} that the positive measure of the  
K$\ddot{a}$llen-Lehmann spectral representation has 
already implied a non-negative anomalous dimension, 
namely $\eta\ge 0$. However, this is not the 
case when the scalar field couples to a gauge field, 
say the case a superconductor is placed in a magnetic 
field. In this case, a negative anomalous dimension 
for the scalar field is unanimously established both 
from perturbative calculations and Monte Carlo 
simulations performed on a  lattice (see Ref. 
\cite{fla} and references therein). More recent works 
showed that a negative anomalous dimension actually 
signals that the system experiences a first order phase
transition through the Lifshitz scenario\cite{Fla-Hag}. 
On the other hand, the proof\cite{parasi,CG,fla} of 
the non-negative anomalous dimension is not applicable 
to systems that are either non-local or have massless 
gauge field degrees of freedom \cite{Fla-Hag}. Thus 
the negative $\eta$ appeared in our NCLG theory does 
not mean that the theory is ill-defined; it is due to
the non-locality of the NCLG. On the contrary, as we 
will show in this section, such a negative $\eta$ 
implied that NCLG theory will also experience a first 
order phase transition to a modulated phase through 
the Lifshitz scenario. Such a phase transition is a 
manifestation of the UV/IR mixing in the NCLG theory.

To see what is going on, we Fourier transform the 
correlation function (\ref{corrscaling}), resulting 
in the asymptotic behavior of Green's function 
(susceptibility)
\begin{eqnarray}
\label{momGree}
G(k)&\sim& k^{2-D-\varepsilon(1+\theta^2/72)}, \\ \nonumber
    &=& k^{-2-\varepsilon\theta^2/72}.
\end{eqnarray}
With the critical value $\theta_c=12/\sqrt{\varepsilon}$, 
the asymptotic behavior of $G(k)$ becomes
\begin{equation}
\label{AsyG}
G(k)\sim\frac{1}{k^4}. 
\end{equation}
This result is remarkable. While we are originally dealing 
with a system with kinetic energy  $\sim k^2$, the result 
(\ref{AsyG}) tells us that, due to the quantum fluctuations, 
the dispersion law is actually modified to be dominated 
by $k^4$ in the long wavelength limit. Normally as long as 
the $k^2$ term has a positive coefficient, it maintains the 
system stable, and dominates the low energy dynamical 
behavior. Therefore, the scenario consistent with (\ref{AsyG}) 
at $\theta_c$ should be that the $k^2$-part of the kinetic 
energy vanishes exactly at the critical $\theta_c$, so that 
the long wavelength behavior of the system at $\theta_c$ is 
dominated by the $k^4$-term. Thus we are led to the 
following effective action in the low energy limit
\begin{equation}
\label{low}
S=-\frac{1}{2}\int^{\Lambda}\frac{d^{D}k}{(2\pi)^D}
[(1-au\theta^2)k^2+bk^4+m^2]\phi(-k)\phi(k)-\frac{1}{4!}
\int^{\Lambda}u(4321)\phi(4)\phi(3)\phi(2)\phi(1),
\end{equation}
where coefficients $a,b$ are positive numbers to 
maintain the stability of the system. In the four 
and above dimensions, the RG analysis in previous 
sections tells us the fixed point action is Gaussian, 
namely $u^{*}=0$; in this case, we don't need to worry 
about the higher derivative terms at all, since the 
$k^2$-term is positive. However, in $4-\varepsilon$ 
dimension, we have a nontrivial NCWF fixed point, which 
makes $u^{*}=16\pi^2\varepsilon/3$ nonzero. In this 
case, the value of $a$ is determined to be $3/2304\pi^2$ 
within the accuracy of the $\varepsilon$-expansion. It 
is necessary to keep track of the $k^4$ term, because 
the sign of the $k^2$ term may possibly change at large 
$\theta$. Actually, if we go back to eq. (\ref{shellinteg})
, expand Besell function $J_0(x)$
to higher order, and take the integral over angular
variable, one can immediately see that we indeed have
positive coefficient for $k^4$ term.
 (In contrast, in the ordinary $\varphi^4$
theory, the sign of the $k^4$ term is negative even up to
two loops\cite{drouffe}.) The value of the coefficients are not 
really important: As we will show below, for the physical 
picture we are going to extract to work, the crucial 
thing is just the sign of $a$ and $b$ in eq. (\ref{low}).

The commutative counterpart of action (\ref{low}) has 
been studied in the context of phase transition\cite{domb}. 
The most interesting thing is that it allows the existence 
of a multi-critical point, more specifically, the so-called 
Lifshitz point in the phase diagram of the system. A 
Lifshitz point is defined as {\em a special tri-critical
point in the phase diagram, at which the coefficient of 
$k^2$ vanishes and that of $k^4$ is positive to maintain 
the stability of systems}\cite{Horn}. Therefore, the value 
$\theta_c=12/\sqrt{\varepsilon}$ in $4-\varepsilon$ 
dimensions discussed in the previous section is nothing but 
a Lifshitz point. At this point the correlation function 
(\ref{corrscaling}) qualitatively changes its asymptotic 
behavior, a signal of a phase transition. This phase 
transition is first-order, because it is not related to 
a change of symmetry, which is calibrated by parameter 
$m^2$ in Landau-Ginsburg theory. 

It is easy to see that before the parameter $\theta$ 
reaches its critical value $\theta_c$ from below, the 
$k^2$ term has a positive coefficient, thus the system 
does not suffer any instability, and the only possible 
phase transition is the second order one that occurs 
at $m^2=0$. The ordered phase is uniform. When the 
parameter $\theta$ exceeds $\theta_c$, the $k^2$ term 
has a negative coefficient and the system suffers an 
instability near $k=0$ in momentum space. Fortunately, 
we have the $k^4$ term with a positive coefficient, 
which can help stabilize the system. As a consequence, 
the system develops a soft mode which is associated 
with the minimum of kinetic energy. This minimum of 
kinetic energy occurs at
\begin{equation}
\label{minik}
k_c=\sqrt{\frac{au\theta^2-1}{2b}},
\end{equation}
which corresponds to a minimal kinetic energy
\begin{equation}
\label{minimE}
E_{k}^{min}=m^2-\frac{(1-au\theta^2)^2}{4b}.
\end{equation}
The appearance of this soft mode also induces
a shift in the critical temperature for the 
second order phase transition from $m^2=0$ to 
\begin{equation}
\label{shiftinT}
m^2=\frac{(1-au\theta^2)^2}{4b}.
\end{equation} 
As usual in the Landau-Ginsburg theory, we take 
$m^2=a_2(T-T_c^{(0)})$, then the shifted critical 
temperature is given by
\begin{equation}
\label{shiftedT}
T_c=T_c^{(0)}+\frac{(1-au\theta^2)^2}{4a_2b}.
\end{equation}
Because the energy is minimized at the nonzero 
momentum $k_0$, the ordering below $\tilde{T}_c$ 
for $\theta>\theta_c$ is characterized by the 
following modulating order parameter
\begin{equation}
\label{ordering}
\phi(r)=\phi_0\cos({\bf p}\cdot{\bf r}+\varphi_0),
\end{equation}
where $\phi_0$ is the amplitude and $|{\bf p}|=k_c$ 
the wavevector of this modulating phase, and 
$\varphi_0$ an initial phase. Therefore, there are 
two ordering phases (uniform phase at $\theta<\theta_c$ 
and a modulated phase at $\theta>\theta_c$) in NCLG. 
The existence of a modulated phase (\ref{ordering}) 
is due to the strong quantum fluctuations at large 
noncommutativity, i.e.,  strong quantum fluctuations 
violates the uniform ordering phase and eventually 
the system enters a finite wavelength modulated phase 
to get stabilized.

Before we discuss the property around the Lifshitz 
point, we would like to point out one basic difference 
between our noncommutative modulated phase (NCMP) and 
the ordinary modulated phase (OMP). As pointed out in 
Ref. \cite{jabari}, the parity of a field theory lives 
on NC space is automatically violated. Thus NCMP is 
also a parity violated phase, so it is different 
from ordinary modulated phase. 

To describe the critical behavior in the vicinity of 
the Lifshitz point, we need to introduce additional 
critical exponents. In the NCMP, the dominant momentum 
dependence is $k^4$, implying that we need a new 
exponent $\nu_m$ to characterize the phase transition
between NCMP and the disordered phase ($T>T_c$). A 
simple dimensional analysis gives the asymptotic
of the correlation length $\xi$ as
\begin{equation}
\label{meannu}
\xi\sim m^{-\frac{1}{2}}\sim(T-T_c)^{-\frac{1}{4}}.
\end{equation}
Therefore, the mean field critical component $\nu_m$ is 
\begin{equation}
\label{nu}
\nu_m=\frac{1}{4}.
\end{equation}
Here we only extract the mean field exponent; to go 
beyond this result, it is a folklore in statistical 
physics that we have to invoke RG again. We expect
that the result would be similar to the case from 
the uniform ordered phase to the disordered phase;
namely, after running the RG, $\nu_m$ also picks up 
an $\varepsilon$-dependent correction. What we would 
like to stress is that $\nu_m$ is different from 
$\nu=1/2$ for the transition from the uniform ordering 
phase to the disordered phase even at the mean field 
level. In general, when Lifshitz point is approached, 
there will be a crossover from $\nu_m=1/4$ to $\nu=1/2$. 
This point can also be seen from the RG analysis in 
Sec. V that the anomalous dimension $\eta$ increases 
with noncommutativity and makes the susceptibility 
$1/k^2$ cross over to $1/k^4$ at the critical point.

To further characterize how the phase boundary 
between the modulated and the uniformly ordered 
phases is approached, we introduce a new critical
exponent $\beta_k$. Approaching the Lifshitz point 
on the NCMP segment of the second order phase 
transition line, we expect that the wave vector 
${\bf p}$ of NCMP will be related to the 
dimensionless variable $\theta$ by
\begin{equation}
\label{betak}
p\sim |\theta-\theta_c|^{\beta_k} 
\hspace{1.0cm} (\theta>\theta_c).
\end{equation}
By noting that the coefficient $b$ is not singular 
at $\theta_c$, and expanding (\ref{minik}) around 
$\theta_c$, we get
\begin{equation}
\label{exp}
p\sim |\theta-\theta_c|^{\beta_k}
\sim|\theta-\theta_c|^{1/2}.
\end{equation}
Therefore, at mean field level, we have
\begin{equation}
\label{betakmean}
\beta_k=\frac{1}{2}.
\end{equation}
Of course, higher order fluctuations may well
alter this value and give rise to singularities 
in the shape of the phase boundaries at the 
Lifshitz point. Similar behavior was discussed 
for bi-critical points in Ref. \cite{fisher-nelson}.

In passing, several remarks are in order. (1) 
In ordinary Landau-Ginsburg theory, we do not have 
such a first order phase transition in the low 
temperature phase even if we include quantum 
fluctuations up to two loops, because there 
quantum fluctuations can not induce a positive 
coefficient for $k^4$, thus the system can not 
stabilize at finite momentum\cite{drouffe}; (2) 
the standard way to study the Lifshitz point is 
to consider the commutative version of eq. 
(\ref{low}) as the starting point with the desired  
coefficient. Namely, the higher order derivative 
terms in quadratic action have already been 
included at the tree level. Meanwhile, in the 
NCLG theory, we do not have such a tree level 
higher order derivative terms in the quadratic 
action. Rather we have infinitely many higher 
order derivative terms in the interaction part 
which are hidden in the star product. Upon 
expanding the star product, each order of the 
expansion involves four field variables and 
their higher order derivatives. Therefore, 
intuitively it is the contraction of two fields
with higher order derivatives and the distinct
UV/IR mixing of NCFT that make the desired 
higher order derivatives terms for the quadratic 
action possible. After recognizing this feature 
of the star product, it seems that the appearance 
of the Lifshitz point and the first order phase 
transition in the NCLG is very natural.

\section{Conclusions and discussions}

In this paper, we have presented a Wilsonian RG 
analysis for noncommutative landau-Ginsburg (NCLG)
theory in arbitrary dimensions. The key point of 
this paper is that to classify the relevance 
or irrelevance of interaction operators in a way 
consistent with the intrinsic constraints from 
noncommutative (NC) geometry. More precisely,
we propose {\it not to apply RG transformations 
to the Moyal factor} in the interaction functions, 
though they are momentum dependent. The reasoning
behind this proposal is that NC parameters, though 
dimensionful, are not renormalized, so that they do 
not transform under RG transformations. Therefore 
the Moyal factor coming from the star product 
structure that coherently organizes infinitely many 
higher order derivative terms, is not renormalized.  
The consistency of such a classification has been 
checked at one loop. With the star products 
preserved at one loop, RG analysis can be done 
readily. 


For the NCLG model, the upper critical dimension 
remains to be four, the same as its commutative cousin. 
In a dimension higher than four, there is no nontrivial 
infrared fixed point, and the low energy behavior of 
the theory is mainly controlled by the Gaussian fixed 
point. Thus the noncommutativity effects do not show 
up in $D>4$ dimensions. Exactly in $D=4$, although we 
still do not have a nontrivial fixed point, we argue 
that noncommutativity effects should appear in 
logarithmic corrections to the scaling behaviors. 
In less than four dimension, by invoking a dimensional 
reduction scheme, we developed an $\varepsilon$-expansion 
in NCLG, a nontrivial stable fixed point appears besides 
the unstable Gaussian fixed point, since the star product 
will appear in the fixed point action, we call it as a 
noncommutative Wilson-Fisher (NCWF) fixed point in 
contrast to WF fixed point in the ordinary space. We 
emphasize that this is a new fixed point in the sense 
that we need at least one additional NC parameter
$\theta=\Theta\Lambda^2$, to characterize it, besides 
the usual mass parameter $r=m^2/\Lambda^2$, and 
interaction strength $u=g\Lambda^{D-4}$.

A new manifestation of the UV/IR mixing in the NC 
field theories is uncovered in $D<4$ dimension: 
The one loop self-energy is no longer merely a 
constant that renormalizes the mass term. It also 
contains a momentum dependent part, which gives 
rise to an anomalous dimension to the field. This
results in a negative critical exponent
$\eta=-\varepsilon\theta^2/72$ near the NCWF 
fixed point. The fact that the ``macroscopic''
critical exponent $\eta$ depends on $\theta$, 
the ``microscopic'' defining parameter for 
NC geometry, is a new manifestation of the 
UV-IR mixing. Naively this might seem to 
indicate a non-universal behavior of the 
exponent. In our opinion, we would rather 
view the NC parameter $\theta$ as one of the 
parameters that characterize the RG fixed point 
or the universality class, in addition to the 
usual $u^{*}$ and $r^{*}$. Therefore, more 
precisely, the NCWF fixed point is not really 
a point; we have the NCWF fixed points, forming 
a fixed-point line labeled by the NC parameter 
$\theta$. In this sense, $\theta$ represents 
a marginal parameter. 

Here we would like to mention that the authors 
of Ref. \cite{italy} have presented a proof of 
renormalizability of the NCLG model based on 
the Wilsonian RG equations. The renormalization
conditions proposed in this paper maintain the 
star product with unrenormalized NC parameters, 
in line with the spirit of our RG 
analysis reported in our short paper 
\cite{chenwu}. However, the authors argued 
later in the same paper that at sufficiently 
low momenta the ordinary Wilsonian RG equations 
with no $\theta$-dependence should be recovered, 
so there should be an intermediate momentum 
region in which somehow the Wilsonian RG 
equations with and without $\theta$-dependence
should overlap. We do not agree with this
opinion. Indeed, for the nonlocal theories in which 
the nonlocality is always of a finite
range, so that there is no UV-IR mixing, 
we do agree that at distances much larger than the 
nonlocality scale, the ordinary Wilsonian RG 
equations should be valid without any trace of 
the nonlocality scale. However, in our opinion,
this argument does not apply to an NC space with 
constant coordinate commutators: Due to 
coordinate-coordinate uncertainty relations, 
the nonlocal effects are {\it not} of a 
finite range in space and we do have UV-IR 
mixing. So it is natural to expect that in such 
NC space, nonlocal effects show up at long 
distances due to UV-IR mixing, and the Wilsonian
RG equations always exhibit $\theta$-dependence 
even at very low momenta. 
 
Moreover, we have shown that for large 
$\theta>\theta_c=12/\sqrt{\varepsilon}$, the 
qualitative behavior of the correlation function is 
changed. In momentum space, this change reflects in
the momentum dependence of the susceptibility $G(k)$ 
that changes from $1/k^2$ to $1/k^4$. This signals a 
momentum space instability, that yields a soft mode 
in the kinetic energy, in which the kinetic energy 
is minimized at non-zero momentum $k_0$. Therefore, 
the NCLG model may exhibit a new modulated order,
which is characterized by the wavelength
$\lambda_0=2\pi/k_0$. A first order phase transition 
from the uniform order to a modulated order at 
$\theta_c$ and $T=T_c$, which we identify as a 
Lifshitz point, is developed for large $\theta$ in 
less than four dimension. 

In contrast to the self-consistent Hartree-Fock 
calculation developed in Ref. \cite{gubser}, our 
RG analysis does not support the stripe phase 
scenario in the critical dimension $D=4$. We 
note that the authors of Ref. \cite{gubser} 
conjectured the possibility of a Lifshitz point 
in the NCLG model. Using our RG analysis, we 
have proved that indeed it does appear in 
$4-\varepsilon$ dimension, and that such a 
Lifshitz point appears exactly when the $k^2$ 
coefficient vanishes while the $k^4$ term in 
kinetic energy is positive to maintain stability. 
Due to the existence of such a modulated phase, 
we need a new critical exponent $\nu_m$ to describe 
the Lifshitz point, in the mean field level, $\nu_m=1/4$. 
Concomitantly, the mean field phase transition
temperature from ordered phase to disordered phase 
is shifted. To further describe the first order 
phase transition between modulated phase and uniform 
ordered phase, we introduced a new critical component 
$\beta_k$ to describe the behavior of $k_c$ near the 
Lifshitz point. Within mean field approximation, 
$\beta_k$ takes value $1/2$.

Finally, several remarks on possible applications 
and implications of our work are in order. First, 
the successful generalization of the Wilsonian RG 
to NC field theory seems to imply that it should be 
possible to generalize field theory RG to NC field 
theory as well. Since the star product structure 
should be intact during renormalization, the 
definition of the beta functions and anomalous 
dimensions should remain the same as their commutative 
counterparts. However, the central issue is how to 
incorporate the nonlocality in NCFT into the generalized 
Callan-Symanzik equations. The simplest and natural 
guess for the latter would be  
\begin{equation}
\label{guess}
[\Lambda\frac{\partial}{\partial\Lambda}+
\beta_u\frac{\partial}{\partial u}-
\frac{N}{2}\gamma_{\phi}]\Gamma^{(N)}(k_i, u, \Lambda)=0,
\end{equation} 
with the Moyal phase factor
\begin{equation} 
\label{phase}
\exp(-\frac{i}{2}\sum_{i<j}k_i\wedge k_j) 
\end{equation}
included in the $\Gamma^{(N)}(k_i, u, \Lambda)$, where 
$u$'s are coupling constants. However, we should admit 
that such a proposal is still very speculative. To 
our current knowledge, how to make sense of field theory 
renormalizability in NCFT (especially in NC Yang-Mills)
to arbitrary orders is still a big question, since the 
limit of pushing cut-off to infinity generates new IR 
singularities\cite{seiberg} due to UV/IR mixing. Though, 
as we have seen, the Wilsonian RG scheme can bypass 
this obstacle by introducing both UV cut-off and IR 
cutoff, the Callan-Symanzik equations contain more 
information than the leading asymptotic behavior near 
the stable fixed point and, therefore, need more 
efforts to establish.  

Secondly, it seems to impossible to introduce operator 
product expansion (OPE) in a NCFT due to its non-locality. 
However, perhaps we should not be too disappointed if 
we recall that in an ordinary field theory there is a 
beautiful relation between one loop Wilsonian RG equations 
and OPE coefficients, namely, OPE coefficients uniquely 
determine the one loop Wilsonian RG equations\cite{cardy}.
If we formally reverse the above logic, it seems that 
there should exist an appropriate modification of the 
OPE, since we already know how to make sense of the RG 
equations for the NCLG model. Of course, by no means it 
would be an easy job. Curiosity is the drive for advances 
in science, NCFT provides us new challenges and also new 
opportunities to understand nonlocal field theories which 
have great potetential applications both in condensed 
matter physics and string theory.

\acknowledgments
One of us, G.H.C., thanks Prof. M. P. A. Fisher for
stimulating discussions.  He also thanks Dr. 
F. Nogueira for discussions on the meaning of 
negative anomalous dimension. This research was 
supported in part by the US National Science Foundation 
under Grants No. PHY-9970701.
    

\newpage
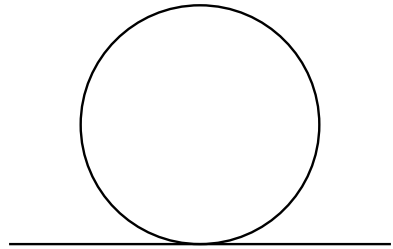
\begin{figure}
\caption{ Tadpole contributions to quadratic action }
\end{figure}
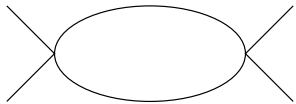
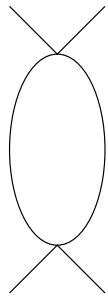
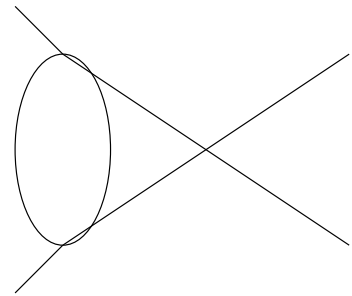
\begin{figure}
\caption{One loop corrections to the quartic vertex}
\end {figure}
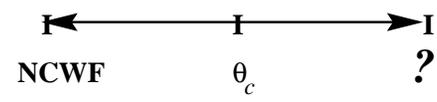
\begin{figure}
\caption{Phase diagram along parameter $\theta$ line}
\end {figure}
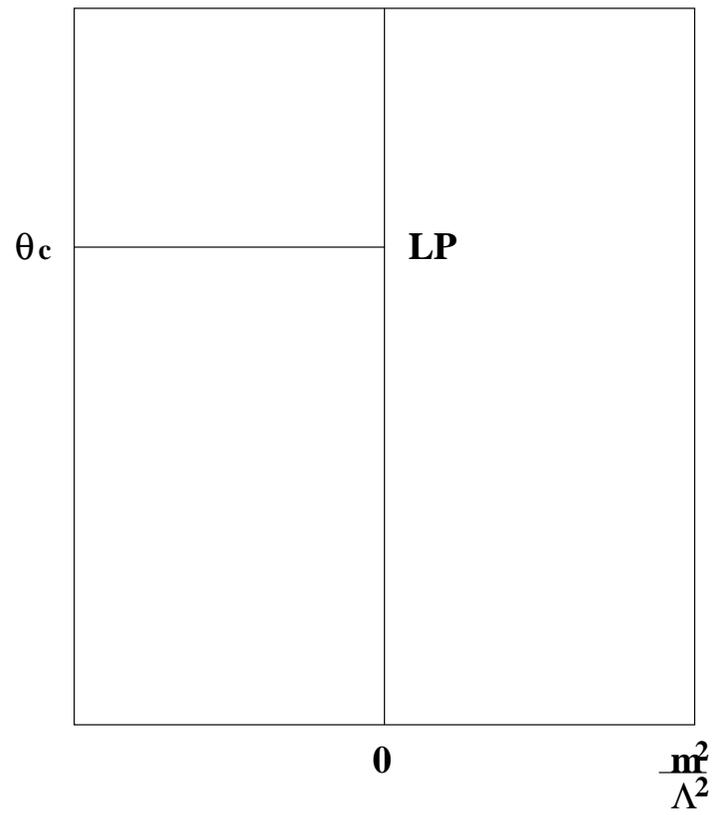
\begin{figure}
\caption{Phase diagram for the Lifshitz scenario. Up:
Lifshitz point in the phase diagram; down: scaling of critical
wave vector $K_c$ }
\end {figure}

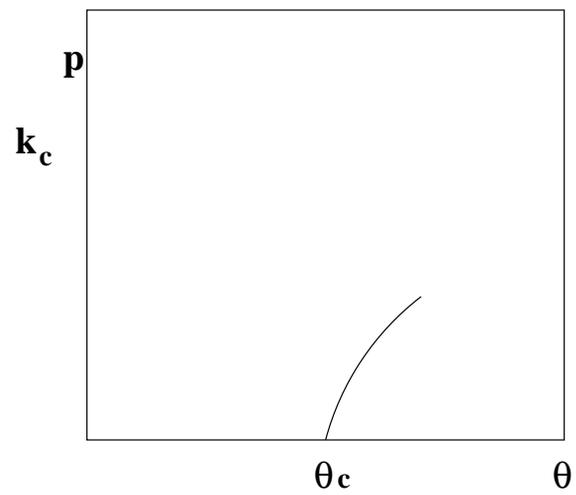
\end{document}